\documentclass{appolb}
\usepackage{epsfig}
\begin{document}
% \eqsec  % uncomment this line to get equations numbered by (sec.num)

\title{Evidence for a narrow N$^*$(1685) resonance
in $\eta$ photoproduction off the nucleon}

\author{V.~Kuznetsov$^{1,2}$, M.V.~Polyakov$^{3,4}$, T.~Boiko$^5$, J.~Jang$^{1} $, A.~Kim$^{1} $,
W.~Kim$^{1,6}$, H. S. Lee$^{1}$, A.~Ni$^{1} $, G.-S.~Yang$^{3}$
\address{$^1$ Kyungpook National University, 702-701, Daegu, Republic of Korea,\\
$^2$Institute for Nuclear Research, 117312, Moscow, Russia,\\
$^3$Institute f\"ur Theoretische Physik II, Ruhr-Universit\"at Bochum,
D - 44780 Bochum, Germany,\\
$^4$ Petersburg Institute for Nuclear Physics, Gatchina, 188300, St. Petersburg, Russia,\\
$^5$ Belarussian State University, 220030, Minsk, Republic of
Belarus,\\
$^6$Daegu Center,  Korea Basic Science Institute, Daegu, 702-701
Republic of Korea }}

\maketitle

\begin{abstract}

Revised analysis of $\Sigma$ beam asymmetry for $\eta$
photoproduction off the free proton from GRAAL is presented. New
analysis reveals a narrow structure near $W\sim 1.685$~GeV. We
describe this structure by the contribution of a narrow resonance
with quantum numbers $P_{11}$, or $P_{13}$, or $D_{13}$. Being
considered together with the recent observations of a bump-like
structure at $W\sim 1.68$ GeV in the quasi-free $\eta$
photoproduction off the neutron, this result provides an evidence
for a narrow ($\Gamma \leq 25$ MeV) N$^{*}(1685)$ resonance.
Properties of this possible new nucleon state, namely the mass,
the narrow width, and the much stronger photocoupling to the
neutron, are similar to those predicted for the non-strange member
of anti-decouplet of exotic baryons.

\end{abstract}


\section{Introduction}

\begin{figure}
\vspace*{-0.4cm}
\centerline{\epsfverbosetrue\epsfxsize=8.5cm\epsfysize=6.5cm\epsfbox{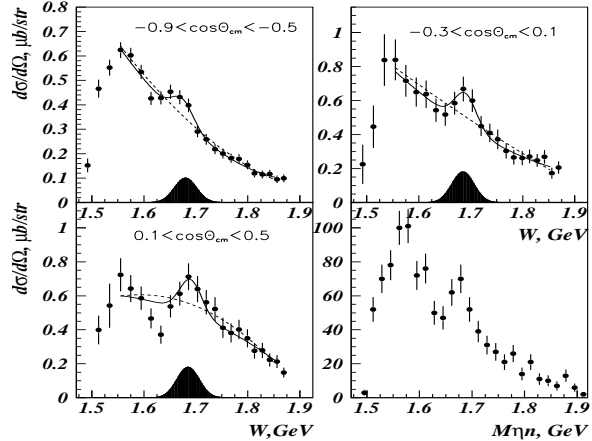}}
\caption{Quasi-free cross sections and $\eta n$ invariant mass spectrum
(low right panel)
for the $\gamma n \to \eta n$ reaction (data from \protect\cite{gra1}).
Solid lines are the fit by the sum of 3-order polynomial and
narrow state. Dashed lines are the fit by 3-order polynomial only.
Dark areas show the simulated signal of a narrow state. }
\label{fig:etan}
\vspace{-0.3cm}
\end{figure}

\begin{figure}
\vspace*{-0.4cm}
\centerline{\epsfverbosetrue\epsfxsize=7.5cm\epsfysize=6.5cm\epsfbox{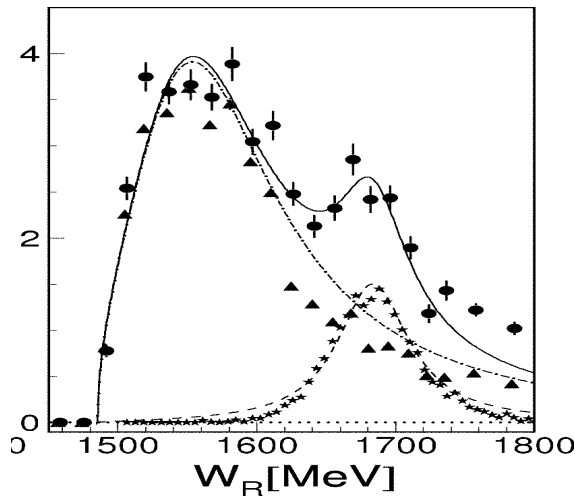}}
%\epsfverbosetrue\epsfxsize=5.5cm\epsfysize=5.9cm\epsfbox{kru_fig4a.eps}
%\includegraphics{as1}% Here is how to import EPS art
\caption{ $M(\eta n)$ spectrum from CBELSA/TAPS \protect\cite{kru}
(filled circle) in comparison with $M(\eta p)$ spectrum (filled
triangles). Stars show the simulated signal of a narrow state with
zero width.} \vspace*{-0.3cm} \label{fig:kru}
\end{figure}

$\eta$ photoproduction off the nucleon is a unique tool to explore
nucleon states with isospin $1/2$. Experimental studies of $\eta$
photoproduction off the proton \cite{etap,gra2,gra3,bon1} resulted
in rich information about low-lying nucleon excitations.
Experiments on $\eta$ photoproduction off the quasi-free neutron
(bound in $^2H$, $^3He$, and $^4He$) until recently were limited
to low photon energies $E_\gamma \leq 820$~MeV
\cite{Krusche:1995zx,HoffmannRothe:1997sv,Hejny:1999iw,Weiss:2001yy,Weiss:2002tn}.
They made it possible to determine the isospin structure of the
$S_{11}(1535)$ resonance \cite{KruSchad}. At higher energies the
rapid rise of the neutron to proton cross section ratio near
$E_\gamma\approx 1$~GeV was observed at GRAAL~\cite{sla02}.
Further measurements at this facility~\cite{Kuznetsov04,gra1}
revealed an interesting phenomenon, a bump-like structure in the
neutron cross section (Fig.~\ref{fig:etan}) near $E_\gamma \sim
1.03$ GeV (the invariant energy  $W\sim 1.68$ GeV). This
observation has been recently confirmed by two other groups:
CBELSA/TAPS~\cite{kru} and LNS-Sendai~\cite{kas}. All three
experiments found a bump in the quasi-free cross-section off the
neutron\footnote{Let us call it as the ``neutron anomaly" because
the quasi-free cross section is affected strongly by the Fermi
motion and by rescattering/final-state interaction. This
observable is more difficult for a theoretical analysis than the
cross section off the free nucleon.}. The width of the bump is
close to that expected for a signal of a narrow resonance smeared
by Fermi motion of the target neutron. In addition, the GRAAL and
CBELSA/TAPS groups observed a narrow peak in the $\eta n$
invariant mass spectrum at $1680-1685$~MeV. The positions of the
peaks are $\sim 1680$~MeV at GRAAL data (low right panel of
Fig.~\ref{fig:etan}) and $\sim 1683$~MeV at CBELSA/TAPS
(Fig.~\ref{fig:kru}). The widths of the peaks (40~MeV in the GRAAL
data and $60\pm 20$~MeV in the CBELSA/TAPS data) are close to the
instrumental resolutions. Such strong peak structure was not
observed in the $\eta$ photoproduction off the proton \cite{etap}.

The anomalous behaviour of the quasi-free neutron cross-section
and the narrow peak the $\eta n$ invariant mass spectrum calls for
a theoretical explanation. A partial-wave analysis of the
quasi-free neutron cross-section is rather complicate because the
target neutron is bound in the deuteron. That is why the search
for this narrow structure (possibly strongly suppressed) in the
$\eta$ photoproduction off the free proton is important. In the
present paper we revise the GRAAL data on $\Sigma$ beam asymmetry
for $\eta$ photoproduction off the free proton. Our goal is to
look for peculiarities near $W\sim 1.68$~GeV in the dependence of
the beam asymmetry on the photon energy.

A simple and concise explanation of the ``neutron anomaly" and the
peak in the $\eta n$ invariant mass is the existence of a narrow
nucleon resonance with much stronger photocoupling to the neutron
than to the proton. Actually, such option was suggested prior the
observation of the `` neutron anomaly" \cite{max,dia1,str}.
Therefore, before the discussion of experimental data, in the
section~2 we provide details on logic and history of this
prediction. Then, in the section~3, we discuss the current
state-of-art in $\eta$ photoproduction off neutron. In the
section~4 we present the revised analysis of the free-proton
$\Sigma$ beam asymmetry for the $\eta$ photoproduction from GRAAL.
In the section~5 our main results and concluisons are summarized.

\section{On predictions of non-strange pentaquark}

If the exotic $S=+1$ pentaquark $\Theta^+$ would exist, this would
imply the existence of a new-type (beyond octet, decouplet and
singlet) flavour multiplet of baryons. The simplest possibility
that is realized in the Chiral Quark Soliton model ($\chi$QSM)
\cite{dia}, is the anti-decouplet of baryons. The anti-decouplet
contains ten baryons. Three of them are explicitly exotic (i.e.
their quantum numbers can not be build out of three quarks only).
The other seven baryons have non-exotic quantum numbers. The
non-strange members of the anti-decouplet are two nucleon states
(two isospin partners): the neutral state ($n^*$) and the
positively charge one ($p^*$). In the $\chi$QSM the spin-parity
quantum numbers of the anti-decouplet members are unambiguously
predicted to be $J^P=\frac 12^+$ \cite{dia}, so that the N$^*$
from the anti-decouplet was predicted to be a $P_{11}$ nucleon
resonance. The idea of the authors of Ref.~\cite{dia} was to
identify the N$^*$ from anti-decouplet with the known
$P_{11}(1710)$ resonance. The choice has been made because of the
following reasons:
\begin{itemize}
\item
Dynamical calculations in the $\chi$QSM gave the mass of N$^*$ in
the range of $1650\div 1750$~MeV.
\item
At the time Particle Data Group \cite{PDG96} reported
the partial decay branchings of $P_{11}(1710)$ consistent with the pattern
predicted for the decays of anti-decouplet: strong coupling to
$\eta N, K\Lambda$ and $\pi \Delta$ channels with  suppression of
$\pi N$ decay mode.
\item
In 1997 the total width of $P_{11}$ was very uncertain and could
accommodate the narrow width of $\leq 40$~MeV predicted by
Ref.~\cite{dia} for the N$^*$ from anti-decouplet.
\end{itemize}
The last point concerns the total width. It was not easy
for the authors of Ref.~\cite{dia} to adopt that
so small width of $\leq 40$~MeV was barely compatible with the
data. The authors thought about existence of a new nucleon resonance in
this mass region, therefore they quoted the result of the Zagreb
group:\\
\noindent {\it However, it should be mentioned that a recent
analysis~\cite{Batinic} suggests that
there might be two nucleon resonances in the region of $\sim 1700$~MeV:
one coupled stronger to pions and another to the $\eta$ meson.}\\
\noindent On other hands, at that time it was hard to believe that
intensive studies of baryon spectroscopy for many years could miss
a relatively light and narrow $\Gamma \leq 40$~MeV nucleon resonance.

First reports \cite{Nakano,Dolgolenko} on the observation of the
exotic $\Theta^+$ pentaquark (begining of 2003) with the mass
close to predicted in $\chi$QSM~\cite{dia,Praszalowicz:2003ik}
rose anew questions about the non-strange member of the
anti-decouplet. It was suggested in Ref.~\cite{max} that
photoproduction of mesons off the neutron can be used as a
benchmark to reveal the anti-decouplet nature of a nucleon
resonance. The transition $\gamma p\to p^*$ of the N$^*$ from the
anti-decouplet should be strongly suppressed relatively the to the
$\gamma n\to n^*$. This suppression is proportional to SU$_{\rm
fl}(3)$ symmetry breaking and can be as large as 1/10 in
scattering amplitudes. The same paper \cite{max} suggested to
probe the anti-decouplet nature of a nucleon resonance by using
$\gamma n\to \eta n $ and $\gamma n\to K\Lambda$ reactions.
Predictions of Ref.~\cite{max} stimulated one of us (V.~K.) to
push forward the study the $\eta$ photoproduction off the neutron
at GRAAL. In 2004 these efforts had led to the observation of the
``neutron anomaly" \cite{Kuznetsov04}. This finding was firstly
taken sceptically by a part of the (former) GRAAL Collaboration
(see, for example, ~\cite{reb}). Nevertheless, after numerous
checks, the result has been published \cite{gra1}. Now it is
confirmed by the CBELSA/TAPS \cite{kru} and LNS-Sendai \cite{kas}
collaborations. The discussion on the ``neutron anomaly" is given
in the next section.

By autumn of 2003 reports on the observation of the exotic
$\Theta^+$ baryon were piling up. At that time it had became clear
that if $\Theta^+$ exists, it is very narrow. Most of these
evaluations were obtained from the re-analysis of $KN$ scattering
data \cite{kn}. Here we skip the discussion of details. An
important point is that the tentative $\Theta^+$ should be very
narrow, having the width in the (sub)MeV range. Obviously, so
small $\Gamma _{\Theta^+}$ is in contrast with the width $\sim
100$~MeV ascribed~\cite{PDG06} to the $P_{11}(1710)$ resonance.
Consequently the existence of a new nucleon resonance with the
mass near $\sim 1700$ MeV was suggested in Refs.~\cite{dia1,str}.
The authors of Ref.~\cite{dia1} used the Gell-Mann--Okubo mass
relations in the presence of mixing, in order to predict the mass
of this new nucleon resonance. As an input for the
Gell-Mann--Okubo  mass formula the authors of Ref.~\cite{dia1}
used the mass of the reported by the NA49 collaboration \cite{alt}
$\Xi^{--}$ baryon. In Ref.~\cite{str}, in order to constrain the
mass of this possible new narrow N$^*$, the modified PWA of $\pi
N$ scattering data was employed. It was found that the easiest way
to accommodate a narrow N$^*$ is to set its mass around $1680$~MeV
and quantum numbers to $P_{11}$ ($J^P=\frac 12^+$). In the same
paper the width of the possible N$^*$ was analysed in the
framework of $\chi$QSM. It was found that the width of new N$^*$
is in range of tens of MeV\footnote{The analysis is rather
uncertain due to large uncertainty in the mixing angle of N$^*$
with ground state nucleon} (most probably below 30~MeV if one
combines the model analysis with modified PWA). Extensive studies
of the decay widths of anti-decouplet baryons in the framework of
$\chi$QSM were performed in Refs.~\cite{michal1,michal2,michal3}.
It was shown that the SU$_{\rm fl}(3)$ symmetry breaking effects
contribute considerably to the partial widths of the non-strange
member of the anti-decouplet. In particular, they suppress the
partial decay $N^*\to \pi N$ whereas provide a small contribution
to the $\eta N$ and $K\Lambda$ decay modes. The phenomenological
analysis of baryon spectrum in the framework of broken flavour
SU$_{\rm fl}(3)$ of Refs.~\cite{guz2} suggests that the width of
non-strange member of the anti-decouplet is below of $50$~MeV.

Recent different $\chi$QCM calculations
\cite{dia5,Lorce,Ledwig:2008rw} of the anti-decouplet widths have
shown that the width of $\Theta^+$ is in (sub)MeV region and the
width of the non-strange partner N$^*$ is in the range
$15-20$~MeV. Detailed discussion of the narrow widths of
pentaquarks in the $\chi$QSM is available in
Ref.~\cite{Diakonov:2006kh}.

\section{$\eta$ Photoproduction off the neutron}

The study of the quasi-free $\gamma n \to \eta n$ reaction at
GRAAL \cite{Kuznetsov04,gra1} (Fig.~\ref{fig:etan}),
CBELSA/TAPS~\cite{kru}(Fig.~\ref{fig:kru}), and LNS-Tohoku
\cite{kas} facilities provided an evidence for a relatively narrow
structure at invariant energy $W\sim 1.68$ GeV. The structure has
been observed as a bump in the quasi-free cross section and in the
$\eta n$ invariant mass spectrum. The width of the bump in the
quasi-free cross section was found to be close to that expected
due to smearing by Fermi motion of the target neutron bound in the
deuteron. A narrow resonance, which would manifest as a peak in
the cross section off the free neutron, would appear in the
quasi-free cross section as a bump of about 50~MeV
width~\cite{gra1} (Fig.~\ref{fig:etan}). The simulated signal of
such resonance (folded with momentum distribution of bound
neutron) with the mass $M\sim 1.68$ GeV and the width
$\Gamma=10$~MeV is shown in Fig.~\ref{fig:etan}. The cross section
is well fitted by the sum of a background and the contribution of
this resonance.

The $\eta n$ invariant mass is almost unaffected by Fermi motion.
The narrow peak in the $\eta n$ invariant spectrum mass cannot
originate from rescattering effects. The widths of the peaks in
the $M(\eta n)$ spectra ($40$ MeV at GRAAL~\cite{gra1} and $~60$
MeV at CBELSA/TAPS~\cite{kru}) are nearly equal to the
instrumental resolutions.

Such bump is not seen in $\eta$ photoproduction off the proton.
The cross section off the proton exhibits only a minor peculiarity
in this mass region~\cite{etap}. Therefore the bump in $\eta$
photoproduction off the neutron may signal a nucleon resonance
with unusual properties: the mass $M\sim 1.68$~GeV, the narrow
width, and the much stronger photocoupling to the neutron than to
the proton.

On the base of the data from Refs.~\cite{Kuznetsov04,gra1}, the
photocoupling of the tentative N$^*$ was estimated in
Ref.~\cite{akps} as $\sqrt{{\rm Br}_{\eta N}} A_{1/2}^n \sim
15\cdot 10^{-3}$~GeV$^{-1/2}$\footnote{Possible theoretical errors
of this analysis are up to a factor of two.}. This value is in
good agreement with $\chi$QCM calculations \cite{yang}. The
influence of a narrow resonance on various observable was
investigated in Ref.~\cite{kim}. It was shown that the inclusion
of a narrow resonance could describe the experimental data.
However, important effects of Fermi motion of the target neutron
were ignored in this publication. The inclusion of such resonance
into the Reggeized version of an isobar model for $\eta$
photoproduction $\eta$-MAID~\cite{maidR} generates a narrow peak
in the cross section off the free neutron. This peak is
transformed into a wider bump similar to experimental observation,
if the Fermi motion is taken into account~\cite{tia1}.

The standard $\eta$-Maid isobar model  \cite{maid} provides an
enhancement in the neutron cross section over the proton one for
$E_\gamma \geq 1$~GeV due to the contribution of $D_{15}(1675)$
resonance. This resonance has stronger coupling to the neutron
than to proton. However it is $\sim 150$~MeV wide. The
contribution of the $D_{15}(1675)$ cannot explain the narrow bump
in the $\eta n$ invariant mass spectrum. Moreover, the standard
$\eta$-Maid \cite{maid} isobar model uses the branching ratio for
the decay $D_{15}(1675) \to \eta N$ of  $0.17$. This value is in
the sharp contrast with the PDG value of $Br_{\eta N}=0.00\pm
0.01$. Also so large branching ratio contradicts the SU$_{\rm
fl}(3)$ analysis of the baryon decays in Ref.~\cite{guz2} which
limits this branching to the range 0.02-0.03.

Alternative explanations of the ``neutron anomaly" was suggested
in Ref.~\cite{skl,aniso}. The authors demonstrated that the bump
in the $\gamma n \to \eta n$ cross section could be explained in
terms of photoexcitation and interference of the known
$S_{11}(1650)$ and $P_{11}(1710)$ (or $S_{11}(1535)$ and
$S_{11}(1650)$) resonances. However, the authors did not discuss
how to explain the narrow bump in $\eta n$ mass spectrum in the
GRAAL \cite{gra1} and the CBELSA/TAPS \cite{kru} data. Anyway, the
generation of a narrow bump in the $\gamma n \to \eta n$ cross
section due to the interference of known resonances requires a
fine tuning of the neutron photocouplings of these resonances,
without changing the proton ones. This implies that the models
~\cite{skl,aniso} predict no any narrow structure in observables
in the proton channel. On contrary, the narrow  N$^*$ would
produce a narrow structure in observables off the proton, even if
its photoexcitation off the proton is suppressed due to the
SU$_{\rm fl}(3)$ symmetry breaking effects. This is a benchmark
test for these models.

Any decisive conclusion about the nature of the anomalous behavior
of the neutron cross section requires a complete partial-wave
analysis. This procedure is sophisticated: model calculations are
usually performed for the free neutron while measured quasi-free
observables are smeared by Fermi motion. The significant influence
of Fermi motion on differential cross sections is shown in
Ref.~\cite{tia1}. Moreover, the quasi-free cross section is
distorted by re-scattering and final-state interaction (FSI).
Those events which originate from re-scattering and FSI, are in
part eliminated in data analysis. Accordingly the measured
quasi-free cross section might be smaller than the calculated
cross section off the free neutron smeared by Fermi motion.

In this sense free-proton data are much more attractive. If
photoexcitation of a nucleon resonance occurs on the neutron its
isospin partner should be produced in the proton channel as well.
However a strong suppression in the proton channel is possible.
For example, the exact SU$_{\rm fl}(3)$ would forbid the
photoexcitation of the non-strange pentaquark from the
anti-decouplet off the proton. Accounting for the SU$_{\rm fl}(3)$
violation leads to the cross section of its photoproduction off
the proton 10-50 times smaller (but not 0) than that off the
neutron~\cite{max,yang}.

\section{$\eta$ Photoproduction off free proton }

$\eta$ photoproduction off the proton below $W\sim 1.7$~GeV is
dominated by photoexcitation of the $S_{11}(1535)$ resonance. This
resonance contributes to the $E_{0}^{+}$ multipole only.
$|E_{0}^{+}|^2$ is the major component of the cross section
\vspace*{-0.2cm}
\begin{equation}
\sigma \sim |E_{0}^{+}|^2+{\rm interference\ terms}
\end{equation}
while other multipoles contribute through the interference
with $E_{0}^+$ or between themselves.
A narrow weakly-photoexcited state with the mass below $1.7$ GeV would
appear in the cross section as a small peak/dip structure
on the slope of the dominating $S_{11}(1535)$ resonance.
In experiment this structure would be in addition smeared by
the resolution of a tagging system (for example, the resolution of
the tagging system at GRAAL is 16 MeV FWHM), and might be masked
due to inappropriate binning.

Polarization observable - the polarized photon beam asymmetry
$\Sigma$ is much less affected by the $S_{11}(1535)$ resonance.
This observable is the measure of azimuthal anisotropy of a
reaction yield relatively the linear polarization of the incoming
photon. In terms of $L\leq 1$ multipoles the expression for the
beam asymmetry does not include the multipole $E_{0}^{+}$:

\begin{eqnarray}
\Sigma(\theta)\sim \frac{3\sin^2\theta}{2}Re(-3|E_1^+|^2+|M_1^+|^2-\\
\nonumber{-2M_1^{-*}(E_1^+-M_1^+)+2E_1^{+*}M_1^+).}
\end{eqnarray}

This observable is mostly governed by the multipoles others than
$E_0^+$ and therefore is much more sensitive to signals of
non-dominant resonances than the cross section. The possible weak
signal of N$^*$ could be amplified in beam-asymmetry data due to
the interference between multipoles.

\begin{figure}
\vspace*{-0.3cm}
\centerline{\epsfverbosetrue\epsfxsize=10.8cm\epsfysize=8.8cm\epsfbox{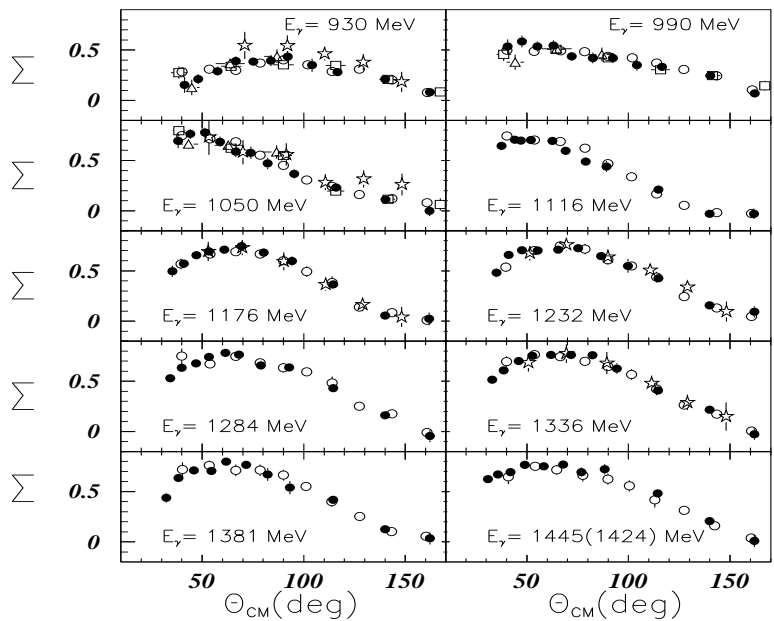}}
\caption{ Published data  for $\Sigma$ beam asymmetry for $\eta$
photoproduction off the free proton. Open triangles and squares
are from \protect\cite{gra2}: squares correspond to the detection
of two photons from $\eta \to 2\gamma$ decays in the BGO ball;
triangles are obtained detecting one photon in the forward shower
wall and the second in the BGO ball. Black circles are the data
from \protect\cite{gra3}. Open circles are from ~\cite{ll}. Stars
are the results from \protect\cite{bon1}.} \vspace*{-0.3cm}
\label{fig:as1}
%\vspace*{-0.3cm}
\end{figure}

For $\eta$ photoproduction off the proton the beam asymmetry
$\Sigma$ was measured at the GRAAL facility\footnote{General
description of the GRAAL facility is available in
~\protect\cite{pi0}.}. First results~\cite{gra2} covered the
energy range from threshold to $1.05$~GeV. Two
statistically-independent and consistent sets of data points were
reported (Fig.~\ref{fig:as1}). These data sets were produced using
two different samples of events:
\begin{itemize}
\item[i)] Events in which two photons from $\eta \to 2\gamma$ decays
were detected in the BGO Ball \cite{bgo}.
\item[ii)] Events in which one of the photons emitted at the angles
$\theta_{lab}\leq 25^{\circ}$ was detected in the forward shower
wall\cite{rw}, and the other in the BGO ball.
\end{itemize}

The second type of events was found to be particularly efficient
at forward angles and energies above $0.9$~GeV.
The contamination of such events at the angles below $50^{\circ}$ reaches 80\%.
The results shown a marked peaking at forward angles and
$E_{\gamma} \sim 1.05$~GeV (see Fig.~\ref{fig:as1}).

An extension to higher energies up to $1.5$ GeV was reported
in \cite{gra3}. Two samples of events were merged and analyzed
together. This made it possible to reduce significantly error bars
at forward angles and to retrieve a maximum in the angular
dependence at $50^{\circ}$ and $E_{\gamma}\sim 1.05$ GeV
(Fig.~\ref{fig:as1})

A new measurement was done at CBELSA/TAPS~\cite{bon1}
using the different technique of the photon-beam polarization,
the coherent bremsstrahlung from a diamond radiator.
The results are in a good agreement with \cite{gra2, gra3} but exhibit
slightly larger error bars (see Fig.~\ref{fig:as1}).

Very recently a new data obtained at the GRAAL facility, was
published in Ref.~\cite{ll}. The data set is based on the full statistics
collected at GRAAL. The results are quite similar to those
presented in Ref.~\cite{gra3} (Fig.~\ref{fig:as1}),
but, despite the triple increase of
statistics, are less accurate at forward angles. The reason is that
the described above second type of events was
excluded from the data analysis without any explanation
of the motivation.

In the previous publications~\cite{gra2,gra3,bon1} the main focus was done
on the angular dependencies of the beam asymmetry. Data points were
produced using relatively narrow angular bins but nearly
60~MeV wide energy bins. Such wide energy bins do not allow
to reveal any narrow peculiarities in the energy dependence of
the beam asymmetry.

An ultimate goal of this work is to produce beam asymmetry data using narrow
bins in energy, in order to reveal in detail
the dependence of the beam asymmetry
on the photon energy in the region of  $E_{\gamma}=0.85 - 1.15$~GeV
(or $W=1.55 - 1.75$~GeV) and to search for a signal of a narrow resonance.

In this paper we present the revised analysis of data collected
at the GRAAL facility in 1998 - 1999. Only two
experimental runs are used in the analysis, in order to avoid
additional (up to $\pm 8$~MeV)  uncertainties in the determination of
the photon energy due to the different adjustments and
calibrations of the GRAAL tagging system in the different run periods.

The data collection was carried out as a sequence of alternate
measurements with two orthogonal linear polarization states of a
photon beam produced through the backscattering of laser light off
6.04~GeV electrons circulating in the storage ring of European
Synchrotron Radiation Facility. The degree of polarization was
dependent on photon energy and varied from 0.5 to 0.85 in the
energy range of $E_{\gamma}=0.85 - 1.15$~GeV.

The procedure of selection of events is similar to
that used in \cite{gra2,gra3}. Two  types of described above events
are considered. The first type of events is identified by means of
the invariant mass of two photons from $\eta \to 2\gamma$ detected in the BGO ball.
The momentum of the $\eta$ meson is reconstructed from photon
energies and angles. The measured parameters of
the recoil proton are compared with ones calculated using
kinematics constrains.
Those events in which one of the photons
is detected in the forward shower wall \cite{rw},
are analyzed in a different way: the initial selection
is done using the missing mass calculated from
the energy of the incoming photon and the measured momentum of the recoil proton.

After that a kinematical fit is applied for both types of events.
The center-of-mass angles of $\eta$ and the $\phi$-angles of the reaction plane
are determined by a $\chi^2$ minimization procedure comparing
the calculated energies and angles in the laboratory system with
the measured ones and their estimated errors.
This procedure provides the most accurate determination the reaction
$\theta$ and $\phi$ angles and allows to reduce the influence of
the detector granularity.
For the second type of events, it also allows the determination the energy
of the photon detected in the forward wall.
After that the events are selected using kinematics constraints
and the value of $\chi^2$.
At the final stage both samples of events are merged
and used together to extract beam asymmetries.

\begin{figure}
\vspace*{0.5cm}
\centerline{\epsfverbosetrue\epsfxsize=11.0cm\epsfysize=11.0cm\epsfbox{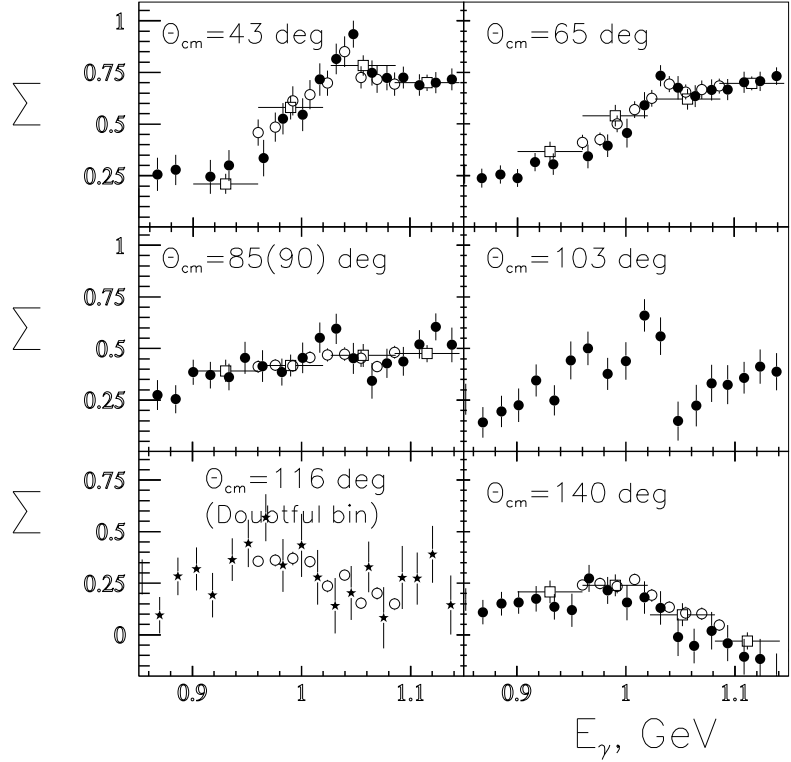}}
\caption{Beam asymmetry $\Sigma$ for the $\eta$ photoproduction
off the free proton obtained here with narrow energy bins (black
circles). Open squares are previous data from
Ref.~\protect\cite{gra3}. Open circles are the data from
Ref.\protect\cite{ll}. Stars are our results at $116^{\circ}$
obtained using the same angular binning as in
Ref.\protect\cite{ll}. } \vspace*{-0.5cm} \label{fig:as2comp}
%\vspace*{-0.3cm}
\end{figure}

The results are shown in Fig.~\ref{fig:as2comp} by filled circles.
They are consistent with the previous data from Ref.~\cite{gra3}.
New data points are obtained using narrow energy bins $\Delta E_{\gamma} \sim 16$~MeV. Angular bins are chosen to be rather wide, about
$20 - 40^{\circ}$, to gain statistics and hence reduce error bars.
At forward angles $\theta_{cm}=43^{\circ}$ and $E_{\gamma}=1.04$~GeV
data points form a sharp peak with the asymmetry in its maximum reaching
values as large as 0.94. The peak becomes less pronounced but clear at $65^{\circ}$.
It is replaced by an oscillating structure at $85^{\circ}$ and at $105^{\circ}$.
At more backward angles the values of asymmetry above $1.05$~GeV drop down almost
to 0 (Fig.~\ref{fig:as1}) while statistical errors grow up.
The peak at forward angles and the oscillating structure at central angles
altogether form an interference pattern which may signal a narrow nucleon resonance.

It is worth to noting that the authors of Ref.~\cite{ll} found
``... no evidence for a narrow $P_{11}(1670)$ state..."
in the beam asymmetry data. In Fig.~\ref{fig:as2comp} our data and
the data from Ref.~\cite{ll} are plotted together. Both data sets are consistent.
Furthermore, at forward angles ($43^{\circ}$) the data sets are nearly statistically independent.
As it was explained above, our results at forward angles are dominated by the events
in which one of the photons from the $\eta \to 2\gamma$ decay is detected
in the forward wall. Such events are not used in ~Ref.\cite{ll}. Their
results  are based on only events in which
both photons are detected in the BGO ball.
Nevertheless both data sets exhibit a sharp peak-like structure.
The major difference is that we observe the oscillating  structure at $103^{\circ}$.
The authors of Ref.~\cite{ll} show the data at $116^{\circ}$ where they do not observe
any structure. However no reliable data can be produced in this ($116^{\circ}$)
angular bin. At the photon energy $1.05$~GeV recoil protons are emitted into a gap
between the forward and the central part of the GRAAL detector where they cannot
be properly detected. The statistics for this particular angular bin
drops considerably due to the low acceptance of the detector. This drop
of statistics is clearly reflected in our large error bars for the $116^\circ$
angular bin (see low left panel of Fig.~\ref{fig:as2comp}). It is
surprising that the authors of Ref.~\cite{ll} have been able to obtain
so small errors in this bin.
It would be helpful if  the authors
of Ref.~\cite{ll} would present their results at the angles near $100^{\circ}$ as well.

\begin{figure}
%\vspace*{0.5cm}
\centerline{\epsfverbosetrue\epsfxsize=11.0cm\epsfysize=11.0cm\epsfbox{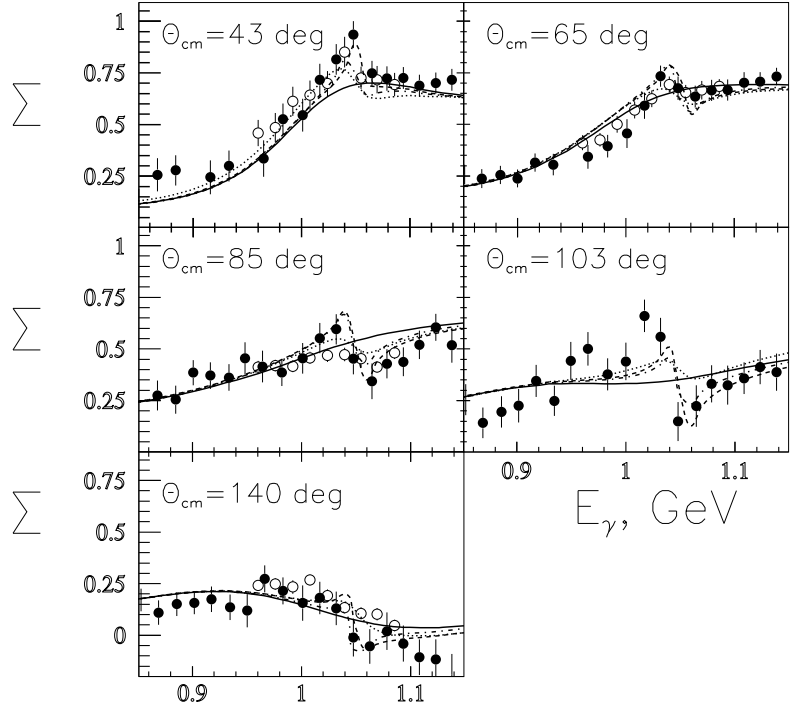}}
\caption{Fit of experimental data (filled circles data obtained in
present analysis, open squares results of
Ref.~\protect\cite{gra3}). Solid lines show our calculations based
on the SAID multipoles only, dotted lines include the $P_{11}$
resonance with the width $\Gamma=19$~MeV; dashed lines are
calculations with the $P_{13}$ ($\Gamma=8$~MeV), while the
dash-dotted lines use the resonance $D_{13}$, also with
$\Gamma=8$~MeV.  Open circles are the data from
Ref.\protect\cite{ll}. } \vspace*{-0.2cm} \label{fig:as2}
\end{figure}

To examine the assumption of a narrow resonance,
we employ the multipoles of the recent E429 solution
of the SAID partial-wave analysis~\cite{str1} for $\eta$ photoproduction as the model
for the smooth part of the observables.
In general, the SAID solution provides good description of the data.
However in the narrow photon-energy interval $E_\gamma=1.015-1.095$
it considerably deviates from the new data (Fig.~\ref{fig:as2}.
The $\chi^2$ value for 24 points in this energy interval at 43, 65, 85, and 103$^{\circ}$
for the SAID solution is $\chi^2/dof=74/24$.

To model this deviation, we add a narrow resonance
(either $S_{11}$, or $P_{11}$, or $P_{13}$, or $D_{13}$)
in the Breit-Wigner form (see e.g. \cite{maid}) to the SAID multipoles.
The contribution of this resonance is parametrized
by the mass, width, photocouplings (multiplied by square of $\eta N$ branching),
and the phase. These parameters are varied, in order to achieve the best
reproduction of experimental data, whereas the SAID multipoles are kept
fixed. We consider the SAID multipoles  as a good approximation
for the non-resonant and/or wide resonances contributions.
The narrow $S_{11}$, $P_{11}$, $P_{13}$, and $D_{13}$ resonances
are tried one by one.
The difference between calculated and experimental
values of the asymmetry $\Sigma$ in the region of
the peak/dip structure (6 points in the energy interval $E_\gamma=1.015-1.095$
in the angular bins of 43, 65, 85,
and 103$^{\circ}$) is used as a criterion for the minimization.

The curves corresponding to the SAID multipoles only are smooth
and do not exhibit any structure (Fig.~\ref{fig:as2}). The
inclusion of either $P_{11}$ or $P_{13}$ or $D_{13}$ resonances
improves the agreement between the data and
the calculations and allows to reproduce the peak/dip structure.
The corresponding values of $\chi^2$ is changed from $\chi^2/dof=74/24$
for the original SAID multipoles
to $\chi^2/dof=56/22$ for the SAID and $P_{11}$, $\chi^2/dof=25/20$ for the
SAID and $P_{13}$, and $\chi^2/dof=39/20$ for the SAID and $D_{13}$ resonances.

The mass of the included resonance is strongly constrained by the
data points. Its values belong to the range of $M_R=1.685 -
1.690$~GeV. The best fit is obtained with the mass
$M_R=1.688$~GeV. However, the extracted mass value includes the
uncertainty of $\pm 5$ MeV which originates from the quality of
the calibration of the GRAAL tagging system, and the uncertainy of
about $\pm 4$~MeV due to the energy binning. Also it may depend on
the basic multipoles used in the fit (in our case SAID
multipoles). That is why at present we quote only the approximate
mass value $M\sim 1.685 $ GeV. The best fit is obtained with the
width of $\Gamma \sim 8$~MeV for $P_{13}$ and $D_{13}$, and
$\Gamma\sim 19$~MeV for $P_{11}$. However, the reasonable
reproduction of the data is achieved up to $\Gamma \leq 25$~MeV.

The $S_{11}$ resonance generates a dip at $43^{\circ}$ in the
entire range of variation of its photocoupling and phase. Its inclusion
does not lead to the improvement of the $\chi^2$.
This indicates that the observed structure most probably
can not be attributed to $S_{11}$.

The curves shown in Fig.~\ref{fig:as2}, corresponds to
\begin{equation}
\sqrt{Br_{\eta N}} A^p_{1/2}\sim 1\cdot 10^{-3}~{\rm GeV}^{-1/2},
\end{equation}
for the $P_{11}$ resonance.
\begin{eqnarray}
\sqrt{Br_{\eta N}} A^p_{1/2}&\sim& -0.3\cdot 10^{-3}{\rm ~GeV}^{-1/2},\\
\sqrt{Br_{\eta N}} A^p_{3/2}&\sim& 1.7\cdot 10^{-3}{\rm ~GeV}^{-1/2},
\end{eqnarray}
for the $P_{13}$ quantum numbers of the resonance. Eventually we obtain
\begin{eqnarray}
\sqrt{Br_{\eta N}} A^p_{1/2}&\sim& -0.1\cdot 10^{-3}{\rm ~GeV}^{-1/2},\\
\sqrt{Br_{\eta N}} A^p_{3/2}&\sim& 0.9\cdot 10^{-3}{\rm ~GeV}^{-1/2},
\end{eqnarray}
for the $D_{13}$ resonance.

The obtained value of $\sqrt{Br_{\eta N}} A^p_{1/2}$ for the
narrow $P_{11}$ resonance is in good agreement with estimates for
the non-strange pentaquark from the antidecouplet performed in
Chiral Quark-Soliton Model~\cite{max,yang}. Comparing the value
with the analogous quantity for the neutron extracted in the
phenomenological analysis of the GRAAL and CBELSA/TAPS data
\cite{akps,tia1}, we obtain the ratio $$A^n_{1/2}/A^p_{1/2}\sim
10-20.$$ This ratio is close to that expected for the non-strange
pentaquark in the Chiral Quark-Soliton model \cite{max,yang}. Such
large ratio of photoproduction amplitudes indicates the strong
suppression of photoexcitation of this resonance off the proton.

\begin{figure}
\vspace*{0.5cm}
\centerline{\epsfverbosetrue\epsfxsize=11.0cm\epsfysize=11.0cm\epsfbox{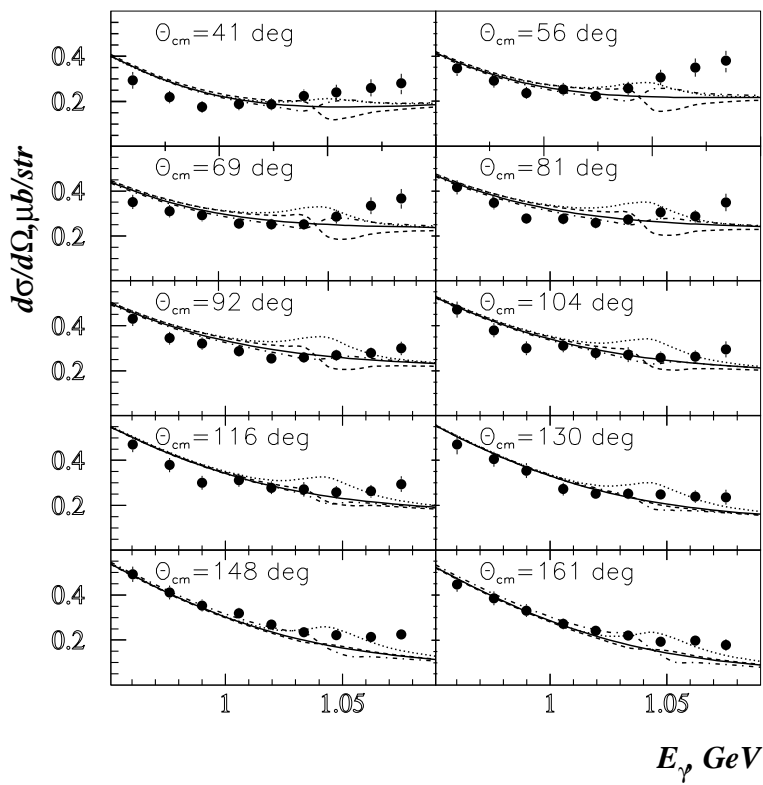}}
\caption{Differential cross section for $\eta$ photoproduction off
the free proton. Black circles are the data from
Ref.\protect\cite{ll}. The legend for curves is the same as in
Fig.~\protect\ref{fig:as2}.} \vspace*{-0.5cm} \label{fig:ds1}
%\vspace*{-0.3cm}
\end{figure}
The calculated differential cross section is shown in
Fig.~\ref{fig:ds1} together with the data from Ref.~\cite{ll}. The
included narrow $P_{13}$ and $D_{13}$ resonances generate only
minor 10-MeV wide structures. The $P_{11}$ generate a ~20-MeV wide
small bump. In our opinion, the quality of the data from
Ref.~\cite{ll} is not enough to reveal such fine peculiarities.
The cross-section data shown in Fig.~\ref{fig:ds1} are smeared due
the resolution of the GRAAL tagging system ($\sigma (E_{\gamma})
=16$ MeV(FWHM)), and by the 16-MeV wide binning. Furthermore, this
data is the compilation from many experimental runs collected at
the GRAAL@ESRF facility during 1998 - 2003. The determination of
the photon energy in each runs includes a systematic shift up to
$\pm 8$ MeV which originates from the adjustment and calibration
of the GRAAL tagging system, and from different operating
conditions of the ESRF. Neither of effort was done in
Ref.~\cite{ll} to reduce this uncertainty. These factors
altogether smooth the data and may hide small peculiarities in the
experimental cross section.

New high-resolution data would be crucial to confirm/close
the existence of this resonance. Recently the CLAS collaboration
reported a relatively narrow structure  at $W\sim 1.7$~GeV in
$\eta$ electroproduction off the proton~\cite{clas_eetap}. This structure
was tentatively explained as a signal of the $P_{11}(1710)$ (or $P_{13}(1720)$)
resonance. To reproduce the data, the width of $P_{11}(1710)$  was set to
$\Gamma=100$ MeV. It would be interesting to fit together
new $\eta$ photo- and electroproduction data
in the region $W=1.62 - 1.72$ GeV.

\section{Summary and discussion}

In summary we report  the evidence for a narrow structure in the
$\Sigma$ beam-asymmetry data for $\eta$ photoproduction off the
free proton. This structure is described by the contribution of a
narrow resonance with the mass $M\sim 1.685 $~GeV and the width
$\Gamma \leq 25$~MeV. Candidates are either the $P_{11}$ or
$P_{13}$ or $D_{13}$ resonances. The mass and width of the
suggested nucleon resonance are consistent with the parameters of
the peak observed in quasi-free cross-section $\eta$
photoproduction off the neutron \cite{gra1,kru,kas}.

The explanation of the bump in the quasi-free neutron cross
sections by the interference effects of known resonances
\cite{skl,aniso} predicts no any narrow structure in the proton
channel. Our new $\Sigma$ beam asymmetry data for $\eta$
photoproduction off the free proton does not support this
asumption.

If to follow the Occam's razor principle the most simple and
concise explanation of  the observations of
Refs.~\cite{gra1,kru,kas} and results of the present paper (see
also \cite{KPB}) is the existence of a narrow nucleon resonance
N$^*(1685)$ with much stronger photocoupling to the neutron than
to the proton. Being a candidate for the non-strange member of the
exotic anti-decouplet, such resonance supports the existence of
the exotic $\Theta^+$ pentaquark. Presently the majority of the
community jumped to the conclusion that $\Theta^+$ does not exist
(see e.g. Ref.~\cite{Close}). The evidences for a new narrow
nucleon resonance -- good candidate for the non-strange
pentaquark-- presented here, encourages the further search for the
$\Theta^+$ baryon. A new approach for this search is suggested in
Ref.~\cite{moskov}. On the other hand, the exact determination of
the quantum numbers of the reported N$^*(1685)$ state is crucial
for the decisive conclusion about its nature. New dedicated
high-resolution experiments are certainly needed for that.

\section*{Acknowledgements}
It is a pleasure to thank the staff of the European Synchrotron
Radiation Facility (Grenoble, France) for stable beam operation
during the experimental run.  We are thankful to Y.~Azimov,
A.~Fix, K.~Goeke, I. Strakovsky, and L.~Tiator for many valuable
discussions. P.~Druck is thanked for support in data processing.
The authors appreciate very much voluntary help with the analysis
and continuous interest, and support to this work of
K.~Hildermann, D.~Ivanov, I.~Jermakowa, M.~Oleinik, and
N.~Sverdlova. We are grateful to B.~Krusche for providing us with
Fig.~2. This work has been supported in part by the Sofja
Kowalewskaja Programme of Alexander von Humboldt Foundation, by
DFG (TR16), and in part by Korean Research Foundation.

%%%%%%%%%%%%%%%%%%%%%%%%%%%%%%%%%%%%%%%%%%%%%%%%%%%%%%


\begin{thebibliography}{99}
\bibitem{etap}
  B.~Krusche {\it et al.},
  %``New Threshold Photoproduction Of Eta Mesons Off The Proton,''
  Phys.\ Rev.\ Lett.\  {\bf 74} (1995) 3736;\\
  %%CITATION = PRLTA,74,3736;%%
    A.~Bock {\it et al.},
  %``Measurement of the target asymmetry of eta and pi0 photoproduction on  the
  %proton,''
  Phys.\ Rev.\ Lett.\  {\bf 81} (1998) 534;\\
  %%CITATION = PRLTA,81,534;%%
  F.~Renard {\it et al.}  [GRAAL Collaboration],
  %``Differential cross-section measurement of eta photoproduction on the
  %proton from threshold to 1100-MeV,''
  Phys.\ Lett.\  B {\bf 528} (2002) 215
  [arXiv:hep-ex/0011098];\\
  %%CITATION = PHLTA,B528,215;%%
  M.~Dugger {\it et al.}  [CLAS Collaboration],
  %``Eta photoproduction on the proton for photon energies from 0.75-GeV to
  %1.95-GeV,''
  Phys.\ Rev.\ Lett.\  {\bf 89} (2002) 222002
  [Erratum-ibid.\  {\bf 89} (2002) 249904];\\
  %%CITATION = PRLTA,89,222002;%%
  V.~Crede {\it et al.}  [CB-ELSA Collaboration],
  %``Photoproduction of eta mesons off protons for 0.75-GeV < E(gamma) <
  %3-GeV,''
  Phys.\ Rev.\ Lett.\  {\bf 94} (2005) 012004
  [arXiv:hep-ex/0311045];\\
  %%CITATION = PRLTA,94,012004;%%
  O.~Bartholomy {\it et al.}  [CB-ELSA Collaboration],
  %``Photoproduction of eta-mesons off protons,''
  Eur.\ Phys.\ J.\  A {\bf 33} (2007) 133;
  %%CITATION = EPHJA,A33,133;%%
\bibitem{gra2}J.~Ajaka {\it et al.},
  %``New measurement of Sigma beam asymmetry for eta meson photoproduction  on
  %the proton,''
  Phys.\ Rev.\ Lett.\  {\bf 81} (1998) 1797.
  %%CITATION = PRLTA,81,1797;%%
\bibitem{gra3} V.~Kuznetsov \textit{et al.}, $\pi N$ NewsLetters \textbf{16}, 160-165(2002),
                Data are available in the SAID data base at \hbox{http://gwdac.phys.gwu.edu}
\bibitem{bon1}
  D.~Elsner {\it et al.}  [CBELSA Collaboration],
  %``Measurement of the beam asymmetry in eta photoproduction off the proton,''
  Eur.\ Phys.\ J.\  A {\bf 33} (2007) 147
  [arXiv:nucl-ex/0702032].
  %%CITATION = EPHJA,A33,147;%%
%\cite{Krusche:1995zx,HoffmannRothe:1997sv,Hejny:1999iw,Weiss:2001yy,Weiss:2002tn}
\bibitem{Krusche:1995zx}
  B.~Krusche {\it et al.},
  %``Near threshold photoproduction of eta mesons from the deuteron,''
  Phys.\ Lett.\  B {\bf 358} (1995) 40.
  %%CITATION = PHLTA,B358,40;%%
\bibitem{HoffmannRothe:1997sv}
  P.~Hoffmann-Rothe {\it et al.},
  %``Break up and coherent photoproduction of eta mesons on the deuteron,''
  Phys.\ Rev.\ Lett.\  {\bf 78} (1997) 4697.
  %%CITATION = PRLTA,78,4697;%%
\bibitem{Hejny:1999iw}
  V.~Hejny {\it et al.},
  %``Near threshold photoproduction of eta mesons from He-4,''
  Eur.\ Phys.\ J.\  A {\bf 6} (1999) 83.
  %%CITATION = EPHJA,A6,83;%%
\bibitem{Weiss:2001yy}
  J.~Weiss {\it et al.},
  %``Exclusive measurement of coherent eta photoproduction from the deuteron,''
  Eur.\ Phys.\ J.\  A {\bf 11} (2001) 371
  [arXiv:nucl-ex/0304009].
  %%CITATION = EPHJA,A11,371;%%
\bibitem{Weiss:2002tn}
  J.~Weiss {\it et al.},
  %``Exclusive measurements of quasi-free eta photoproduction from deuterium,''
  Eur.\ Phys.\ J.\  A {\bf 16} (2003) 275
  [arXiv:nucl-ex/0210003].
  %%CITATION = EPHJA,A16,275;%%
%\cite{Krusche:2003ik}
\bibitem{KruSchad}
  B.~Krusche and S.~Schadmand,
  %``Study of non-strange baryon resonances with meson photoproduction,''
  Prog.\ Part.\ Nucl.\ Phys.\  {\bf 51} (2003) 399
  [arXiv:nucl-ex/0306023].
  %%CITATION = PPNPD,51,399;%%
  \bibitem{sla02}
  V.Kuznetsov et al. [GRAAL Collaboration], Proceedings
of Int.Workshop on the Physics of Excited Nucleons NSTAR2002,
Pittsburgh, USA, Oct.2002, Ed. E.Swanson, World Scientific, p.
267.
\bibitem{Kuznetsov04}
V.~Kuznetsov et al. [GRAAL Collaboration], Proceedings of Workshop
on the Physics of Excited Nucleons NSTAR2004, March, 2004,
Grenoble, France, Eds. J.-P.Bocquet, V.Kuznetsov, D.Rebreyend,
World Scientific, p.197 -203. arXiv:hep-ex/0409032.
%%CITATION = HEP-EX 0409032;%%
\bibitem{gra1} V.~Kuznetsov {\it et al.},
  %``Evidence for a narrow structure at W approx. 1.68-GeV in eta
  %photoproduction off the neutron,''
  Phys.\ Lett.\  B {\bf 647} (2007) 23.
  %%CITATION = PHLTA,B647,23;%%
  [hep-ex/0606065]
\bibitem{kru}
  I.~Jaegle {\it et al.},
  ``Quasi-free photoproduction of eta-mesons of the neutron,''
  arXiv:0804.4841 [nucl-ex].
  %%CITATION = ARXIV:0804.4841;%%
\bibitem{kas}
  F.~Miyahara {\it et al.},
  %``Narrow Resonance At E(Gamma) = 1020-Mev In The D (Gamma, Eta) P N
  %Reaction,''
  Prog.\ Theor.\ Phys.\ Suppl.\  {\bf 168} (2007) 90.
  %%CITATION = PTPSA,168,90;%%
\bibitem{dia}
 D.~Diakonov, V.~Petrov and M.~V.~Polyakov,
  %``Exotic anti-decuplet of baryons: Prediction from chiral solitons,''
  Z.\ Phys.\  A {\bf 359} (1997) 305
  [arXiv:hep-ph/9703373].
  %%CITATION = ZEPYA,A359,305;%%
\bibitem{PDG96}
  R.~M.~Barnett {\it et al.}  [Particle Data Group],
  %``Review of particle physics. Particle Data Group,''
  Phys.\ Rev.\  D {\bf 54} (1996) 1.
  %%CITATION = PHRVA,D54,1;%%
\bibitem{Batinic}
  M.~Batinic, I.~Slaus, A.~Svarc and B.~M.~K.~Nefkens,
  %``pi N $\to$ eta N and eta N $\to$ eta N partial wave T matrices in a
  %coupled, three channel model,''
  Phys.\ Rev.\  C {\bf 51} (1995) 2310
  [Erratum-ibid.\  C {\bf 57} (1998) 1004]
  [arXiv:nucl-th/9501011].
  %%CITATION = PHRVA,C51,2310;%%
\bibitem{PDG06}
  W.~M.~Yao {\it et al.}  [Particle Data Group],
  %``Review of particle physics,''
  J.\ Phys.\ G {\bf 33} (2006) 1.
  %%CITATION = JPHGB,G33,1;%%
\bibitem{Nakano}
  T.~Nakano {\it et al.}  [LEPS Collaboration],
  %``Observation of S = +1 baryon resonance in photo-production from  neutron,''
  Phys.\ Rev.\ Lett.\  {\bf 91} (2003) 012002
  [arXiv:hep-ex/0301020].
  %%CITATION = PRLTA,91,012002;%%
\bibitem{Dolgolenko}
  V.~V.~Barmin {\it et al.}  [DIANA Collaboration],
  %``Observation of a baryon resonance with positive strangeness in K+
  %collisions with Xe nuclei,''
  Phys.\ Atom.\ Nucl.\  {\bf 66} (2003) 1715
  [Yad.\ Fiz.\  {\bf 66} (2003) 1763]
  [arXiv:hep-ex/0304040].
  %%CITATION = YAFIA,66,1763;%%
\bibitem{Praszalowicz:2003ik}
  M.~Praszalowicz,
  %``Pentaquark in the Skyrme model,''
  Phys.\ Lett.\  B {\bf 575} (2003) 234
  [arXiv:hep-ph/0308114].
  %%CITATION = PHLTA,B575,234;%%
\bibitem{max}M.~V.~Polyakov and A.~Rathke,
  %``On photoexcitation of baryon antidecuplet,''
  Eur.\ Phys.\ J.\  A {\bf 18}, 691 (2003)
  [arXiv:hep-ph/0303138].
  %%CITATION = EPHJA,A18,691;%%
\bibitem{reb}D.~Rebreyend, Talk given at the 10th International
   Symposium on Meson-Nucleon Physics and the structure of the Nucleon MENU2004,
   Beijing, August 29 - Sep. 04 2004, China,
   http://www.ihep.ac.cn/menu03/index.html ($\to$ First Circular $\to$ Scientific program).
\bibitem{kn}
  S.~Nussinov,
  %``Some comments on the putative Theta+ (1543) exotic state,''
  arXiv:hep-ph/0307357;\\
  %%CITATION = HEP-PH/0307357;%%
J.~Haidenbauer and G.~Krein,
  %``Influence of a Z(1540)+ resonance on K+ N scattering,''
  Phys.\ Rev.\  C {\bf 68} (2003) 052201
  [arXiv:hep-ph/0309243];\\
  %%CITATION = PHRVA,C68,052201;%%
  R.~N.~Cahn and G.~H.~Trilling,
  %``Experimental limits on the width of the reported Theta(1540)+,''
  Phys.\ Rev.\  D {\bf 69} (2004) 011501
  [arXiv:hep-ph/0311245];\\
  %%CITATION = PHRVA,D69,011501;%%
  R.~A.~Arndt, I.~I.~Strakovsky and R.~L.~Workman,
  %``K+N scattering data and exotic Z+ resonances,''
  Nucl.\ Phys.\  A {\bf 754} (2005) 261
  [arXiv:nucl-th/0311030];\\
  %%CITATION = NUPHA,A754,261;%%
 W.~R.~Gibbs,
  %``The pentaquark in K+ d total cross section data,''
  Phys.\ Rev.\  C {\bf 70} (2004) 045208
  [arXiv:nucl-th/0405024].
  %%CITATION = PHRVA,C70,045208;%%
\bibitem{dia1}D.~Diakonov and V.~Petrov,
  %``Where are the missing members of the baryon antidecuplet?,''
  Phys.\ Rev.\  D {\bf 69} (2004) 094011
  [arXiv:hep-ph/0310212].
  %%CITATION = PHRVA,D69,094011;%%
\bibitem{str} R.~A.~Arndt, Y.~I.~Azimov, M.~V.~Polyakov, I.~I.~Strakovsky and R.~L.~Workman,
  %``Nonstrange and other unitarity partners of the exotic Theta+ baryon,''
  Phys.\ Rev.\  C {\bf 69}, 035208 (2004)
  [arXiv:nucl-th/0312126].
  %%CITATION = PHRVA,C69,035208;%%
\bibitem{alt}
  C.~Alt {\it et al.}  [NA49 Collaboration],
  %``Observation of an exotic S = -2, Q = -2 baryon resonance in proton  proton
  %collisions at the CERN SPS,''
  Phys.\ Rev.\ Lett.\  {\bf 92} (2004) 042003
  [arXiv:hep-ex/0310014].
  %%CITATION = PRLTA,92,042003;%%
%\cite{Ellis:2004uz}
\bibitem{michal1}
  J.~R.~Ellis, M.~Karliner and M.~Praszalowicz,
  %``Chiral-soliton predictions for exotic baryons,''
  JHEP {\bf 0405} (2004) 002
  [arXiv:hep-ph/0401127].
  %%CITATION = JHEPA,0405,002;%%
%\cite{Praszalowicz:2004dn}
\bibitem{michal2}
  M.~Praszalowicz,
  %``SU(3) breaking in decays of exotic baryons,''
  Acta Phys.\ Polon.\  B {\bf 35} (2004) 1625
  [arXiv:hep-ph/0402038].
  %%CITATION = APPOA,B35,1625;%%
\bibitem{michal3}
  M.~Praszalowicz,
  %``SU(3) constraints on cryptoexotic pentaquarks,''
  Annalen Phys.\  {\bf 13} (2004) 709
  [arXiv:hep-ph/0410086].
  %%CITATION = ANPYA,13,709;%%
\bibitem{guz2}
  V.~Guzey and M.~V.~Polyakov,
  %``Mixing and decays of the antidecuplet in the context of approximate  SU(3)
  %symmetry,''
  arXiv:hep-ph/0501010;\\
  %%CITATION = HEP-PH/0501010;%%
V.~Guzey and M.~V.~Polyakov,
  %``SU(3) Systematization Of Baryons: Theoretical Methods And Mixing With The
  %Antidecuplet,''
  Annalen Phys.\  {\bf 13} (2004) 673;
  %%CITATION = ANPYA,13,673;%%
    V.~Guzey and M.~V.~Polyakov,
  %``SU(3) systematization of baryons,''
  arXiv:hep-ph/0512355.
  %%CITATION = HEP-PH/0512355;%%
\bibitem{dia5}
  D.~Diakonov and V.~Petrov,
  %``Estimate of the Theta+ width in the relativistic mean field
  %approximation,''
  Phys.\ Rev.\  D {\bf 72} (2005) 074009
  [arXiv:hep-ph/0505201].
  %%CITATION = PHRVA,D72,074009;%%
\bibitem{Lorce}
  C.~Lorce,
  %``Improvement of the Theta+ width estimation method on the light cone,''
  Phys.\ Rev.\  D {\bf 74} (2006) 054019
  [arXiv:hep-ph/0603231].
  %%CITATION = PHRVA,D74,054019;%%
  %\cite{Ledwig:2008rw}
\bibitem{Ledwig:2008rw}
  T.~Ledwig, H.~C.~Kim and K.~Goeke,
  %``Axial-vector transitions and strong decays of the baryon antidecuplet in
  %the self-consistent SU(3) chiral quark-soliton model,''
  arXiv:0805.4063 [hep-ph].
  %%CITATION = ARXIV:0805.4063;%%
\bibitem{Diakonov:2006kh}
  D.~Diakonov,
  %``The narrow pentaquark,''
  AIP Conf.\ Proc.\  {\bf 892} (2007) 258
  [arXiv:hep-ph/0610166].
  %%CITATION = APCPC,892,258;%%
\bibitem{kim}  K.~S.~Choi, S.~i.~Nam, A.~Hosaka and H.~C.~Kim,
  %``A new N*(1675) resonance in the gamma N --> eta N reaction,''
  Phys.\ Lett.\  B {\bf 636}, 253 (2006)
  [arXiv:hep-ph/0512136];\\
  %%CITATION = PHLTA,B636,253;%%
S.~i.~Nam, K.~S.~Choi, A.~Hosaka and H.~C.~Kim,
  %``A new candidate for non-strangeness pentaquarks: N*(1675),''
  Prog.\ Theor.\ Phys.\ Suppl.\  {\bf 168}, 97 (2007)
  [arXiv:0704.3101 [hep-ph]];\\
  %%CITATION = PTPSA,168,97;%%
    K.~S.~Choi, S.~i.~Nam, A.~Hosaka and H.~C.~Kim,
  %``A new nucleon resonance in eta photoproduction,''
  arXiv:0710.2185 [hep-ph]\\
  %%CITATION = ARXIV:0710.2185;%%
  K.~S.~Choi, S.~i.~Nam, A.~Hosaka and H.~C.~Kim,
  %``eta Photoproduction and N* resonances,''
  arXiv:0707.3854 [hep-ph].
\bibitem{maidR}
  W.~T.~Chiang, S.~N.~Yang, L.~Tiator, M.~Vanderhaeghen and D.~Drechsel,
  %``A reggeized model for eta and eta' photoproduction,''
  Phys.\ Rev.\  C {\bf 68} (2003) 045202
  [arXiv:nucl-th/0212106].
  %%CITATION = PHRVA,C68,045202;%%
\bibitem{tia1} A.~Fix, L.~Tiator and M.~V.~Polyakov,
  %``Photoproduction of eta-mesons on the deuteron above S11(1535) in the
  %presence of a narrow P11(1670) resonance,''
  Eur.\ Phys.\ J.\  A {\bf 32}, 311 (2007)
  [arXiv:nucl-th/0702034].
  %%CITATION = EPHJA,A32,311;%%
\bibitem{skl}  V.~Shklyar, H.~Lenske and U.~Mosel,
  %``eta-photoproduction in the resonance energy region,''
  Phys.\ Lett.\  B {\bf 650}, 172 (2007)
  [arXiv:nucl-th/0611036].
  %%CITATION = PHLTA,B650,172;%%
\bibitem{aniso}
 A.~V.~Anisovich, talk at the NSTAR07, Bonn, September, 2007,\\
 http://nstar2007.uni-bonn.de/talks
\bibitem{maid} W.~T.~Chiang, S.~N.~Yang, L.~Tiator and D.~Drechsel,
  %``An isobar model for eta photo- and electroproduction on the nucleon,''
  Nucl.\ Phys.\  A {\bf 700} (2002) 429
  [arXiv:nucl-th/0110034].
  %%CITATION = NUPHA,A700,429;%%
\bibitem{pi0} V.Bellini \textit{et al.}, \textit{Eur. J. A.}
              \textbf{26}, 299 - 419 (2006).
\bibitem{bgo} F. Ghio {\it et al.}, {\it Nucl. Inst. and Meth. A}
              {\bf 404}, 71, 1998.
\bibitem{rw}  V. Kouznetsov {\it et al.}, {\it Nucl. Inst. and
              Meth. A} {\bf 487}, 128, 2002.
\bibitem{str1} R.~A.~Arndt, W.~J.~Briscoe, I.~I.~Strakovsky, and
    R.~L.~Workman, in progress, \hbox{http://gwdac.phys.gwu.edu}.
\bibitem{str2} SAID multipoles can be obtained
               via ssh said@gwdac.phys.gwu.edu.
\bibitem{ll}   O.~Bartalini \textit{et al.}, \textit{Eur. Phys. J.} A\textbf{33}, 169 (2007), [nucl-ex:0707.1385].
\bibitem{yang}
  H.~C.~Kim, M.~Polyakov, M.~Praszalowicz, G.~S.~Yang and K.~Goeke,
  %``Exotic and nonexotic magnetic transitions in the context of the SELEX  and
  %GRAAL experiments,''
  Phys.\ Rev.\  D {\bf 71}, 094023 (2005)
  [arXiv:hep-ph/0503237].
\bibitem{akps}
  Y.~Azimov, V.~Kuznetsov, M.~V.~Polyakov and I.~Strakovsky,
  %``Extraction of radiative decay width for the non-strange partner of
  %Theta+,''
  Eur.\ Phys.\ J.\  A {\bf 25}, 325 (2005)
  [arXiv:hep-ph/0506236].
\bibitem{KPB}
  V.~Kuznetsov {\it et al.},
  %``Photoproduction off the nucleon revisited: Evidence for a narrow N(1688)
  %resonance?,''
  arXiv:0801.0778 [hep-ex];\\
  %%CITATION = ARXIV:0801.0778;%%
  V.~Kuznetsov, M.~Polyakov, T.~Boiko, J.~Jang, A.~Kim, W.~Kim and A.~Ni,
  %``eta photoproduction on the proton revisited: Evidence for a narrow N(1685)
  %resonance?,''
  arXiv:hep-ex/0703003.
  %%CITATION = HEP-EX/0703003;%%
\bibitem{clas_eetap}
  H.~Densili {\it et al.} [CLAS Collaboration],
  %``Q*2 dependence of the S11(1535) photocoupling and ...,''
  Phys.\ Rev.\ C {\bf76}, 025211 (2007),
  [arXiv:hep-ph/0612150].
\bibitem{Close}
  F.~Close,
%  {\it ``Vanishing pentaquarks,''}
  Nature {\bf 435} (2005) 287;\\
  %%CITATION = NATUA,435,287;%%
   R.~L.~Jaffe,
 % {\it ``Life and death among the hadrons,''}
  AIP Conf.\ Proc.\  {\bf 792} (2005) 97.
  %%CITATION = APCPC,792,97;%%
\bibitem{moskov}
  M.~Amarian, D.~Diakonov and M.~V.~Polyakov,
  %``To see the exotic Theta^+ baryon from interference,''
  arXiv:hep-ph/0612150.
  %%CITATION = HEP-PH/0612150;%%
\end{thebibliography}
\end{document}